\documentclass[fleqn,usenatbib]{mnras}
\usepackage{showyourwork}

\usepackage{newtxtext,newtxmath}

\usepackage[T1]{fontenc}

\DeclareRobustCommand{\VAN}[3]{#2}
\let\VANthebibliography\thebibliography
\def\thebibliography{\DeclareRobustCommand{\VAN}[3]{##3}\VANthebibliography}

\usepackage{graphicx}	
\usepackage{amsmath}	

\usepackage{color}
\usepackage{hyperref}
\hypersetup{%
        colorlinks,
        breaklinks=true,
        plainpages=false,%
        citecolor=[rgb]{0,0.20,0.45},
        linkcolor=[rgb]{0,0.20,0.45},
        urlcolor=[rgb]{0,0.20,0.45},
        bookmarksopen=true,%
        bookmarksnumbered=false,%
        bookmarksdepth=5%
}

\newcommand{\kms}{km s$^{-1}$}
\newcommand{\ms}{m s$^{-1}$}
\newcommand{\bpb}{$\beta$~Pictoris~b}
\newcommand{\mj}{$\mathrm{M_{Jup}}$}
\newcommand{\rj}{$\mathrm{R_{Jup}}$}

\title[\bpb{} exosatellite limits]{Upper limits on exosatellites around \bpb{}}

\author[M. A. Kenworthy et al.]{Matthew A. Kenworthy,$^{1}$\thanks{E-mail: kenworthy@strw.leidenuniv.nl}
Rico Landman,$^{1}$
Andrew Vanderburg,$^{2,3}$
Joseph E. Rodriguez,$^{4}$\newauthor
Jayne L. Birkby,$^{5}$
Isabella Macias,$^{6,7}$
Dar{\'i}o Gonz{\'a}lez Picos,$^{1}$
Sydney A. Jenkins,$^{4}$
Elina Kleisioti,$^{1,8}$\newauthor
Tomas Stolker,$^{1}$
Ioannis Koutalios$^{1}$
\\
$^{1}$Leiden Observatory, Leiden University, Postbus 9513, 2300 RA Leiden, The Netherlands\\
$^{2}$Center for Astrophysics | Harvard \& Smithsonian, 60 Garden Street, Cambridge, MA 02138, USA\\
$^{3}$Department of Physics and Kavli Institute for Astrophysics and Space Research, Massachusetts Institute of Technology, Cambridge, MA 02139, USA\\
$^{4}$Center for Data Intensive and Time Domain Astronomy, Department of Physics and Astronomy, Michigan State University, East Lansing, MI 48824, USA\\
$^{5}$Astrophysics, University of Oxford, Denys Wilkinson Building, Keble Road, Oxford, OX1 3RH, UK\\
$^{6}$Department of Earth, Atmospheric, and Planetary Sciences, Massachusetts Institute of Technology, 77 Massachusetts Avenue, Cambridge, MA 02139, USA\\
$^{7}$Department of Astronomy, University of Florida, Bryant Space Science Center, Stadium Road, Gainesville, FL 32611, USA\\
$^{8}$Faculty of Aerospace Engineering, Delft University of Technology, 2629 HS, Delft, The Netherlands
}

\date{Accepted 2026 May 29. Received 2026 May 16; in original form 2026 February 13}

\pubyear{2026}

\begin{document}
\pagerange{?--?}
\maketitle

\begin{abstract}
   \bpb{} is one of the closest known directly-imaged gas giant exoplanets with an orbit that is almost edge-on to our line of sight, making it an ideal target for radial velocity monitoring to search for massive exomoons.
We measure the radial velocity of \bpb{} over several epochs between October 2024 and March 2025 by using the cross-correlation of a template spectrum with absorption lines in the planet's atmosphere, giving a mean precision of 160 \ms{}.
The resultant set of radial velocities is analysed with a periodogram to search for candidate RV signals indicating a massive exomoon.  
Although we do not detect an exomoon signal in our data, our detection limits for a single moon are 80 Earth masses at $P=1$~day and 1 Jupiter at $P=200$~days, comparable to RV exomoon searches around other substellar companions.
The RV limit is comparable with the astrometric exomoon limit at a period of 7 days and a mass of 150~$M_\oplus$, where for longer periods the astrometric searches have lower mass limits.
With an additional observing season, CRIRES+ can detect a planet/moon mass ratio of $10^{-3}$ ($4M_\oplus$) with a period of up to one day, and can detect a Neptune-mass moon at hundreds of Jupiter radii.
\end{abstract}



\section{Introduction}

Moons are seen around all of the gas giant planets in our Solar system, and their final configuration reflects both the composition of the planet and its formation process \citep[e.g.][]{Deienno11}.
Such satellites, therefore, must exist around other gas giant planets in the Galaxy - but while there have been some tantalizing candidates detected, none have been confirmed to date.
Exomoons may increase the number of Habitable Zones in a planetary system - their temperature may be raised by the flux of energy from their parent planet \citep{Heller13} or by tidal heating.
Since our current knowledge of moons comes entirely from solar system satellites, an exomoon discovery could similarly boost our understanding of the diversity that we see in exoplanetary systems \citep[e.g.][]{Batalha14}.

To understand the range of possible exomoon properties, it is useful to consider the three primary formation pathways: accretion from a circumplanetary disk \citep{Kane13}, gravitational capture of interplanetary objects \citep[e.g. suggested for Triton by Neptune; ][]{Agnor06}, and collisions between objects.
Moons forming from a circumplanetary disk (CPD) are expected to be relatively small compared to their host planet \citep[masses less than $10^{-5}$ that of the planet;][]{Canup02}.
While this formation pathway is highly efficient in the solar system, moons like Triton likely started off as an interplanetary planetesimal and was captured by Neptune some time later \citep{Agnor06}, resulting in Triton's anomalously large mass ratio with respect to Neptune ($2\times10^{-4}$). 
%

Several methods have been used to search for exomoons, with several candidates proposed; for an overview see the Review in \citet{2024arXiv240113293T}.
Gravitational microlensing by a foreground source with proper motion causes achromatic magnification in the resultant light curve of a distant background source.
This led to \citet{Bennett14} showing the microlensing event MOA-2011-BLG-262 was consistent with a free-floating planet/moon system closer to the Solar system than the background source, although a more distant stellar mass primary/brown dwarf system could produce the same observations.
Model degeneracies also led to \citet{Hwang18} identifying the multi-object microlensing event OGLE-2015-BLG-1459L as due to one object lensing three background sources and not a foreground star/planet/moon lensing system. 

Directly imaging the thermal emission from tidally heated exomoons (THEMs) is a possibility for very close exoplanet systems \citep{Peters13} such as the exoplanet Epsilon~Eridani~b \citep{Kleisioti23}.
Tidally locked THEMs can also be detected on high inclination orbits around their parent planet; any bright spot of thermal emission caused by volcanic activity will result in photometric modulation of the light curve \citep{Kleisioti24}.
Volcanic exomoons can also produce an exosphere that is detected as spectral absorption lines of neutral Sodium and Potassium seen before, during and after the transits of hot Jupiters, as seen for WASP-49 b \citep{Oza19}.

Exomoons may transit their parent exoplanet, and searches around free floating \citep{Limbach21,Limbach24b,Wilson25} and bound \citep{2025arXiv251115317K,2026arXiv260405235P} self-luminous exoplanets show plausible eclipses, although the variability of the planets themselves can confound this signal.

Searching for exomoons as small as Ganymede is possible through the impact an exomoon has on the transit of its parent planet across the stellar disk.
This can be seen in both transit timing variations and transit duration variations \citep{Kipping09,Kipping09a}, and were the main methods behind the Hunt for Exomoons with Kepler \citep[HEK; ][]{Kipping12,Kipping13,Kipping22}, but to date no compelling candidates have been found around gas giants with orbital periods less than 50 days \citep{Kipping15}.
Since it is expected that the planets forming in the 5--20 au regime have attendant moons, it is suspected that the process of their migration caused the moons to be scattered away from the system \citep{Dobos21}.
The question of whether wider orbit planets host such satellites arises: two exomoon candidates have been identified through timing variation and possible detection of their transits, \mbox{Kepler~1708~b-i} and \mbox{Kepler~1625~b-i} \citep{Teachey18,Kipping22,Teachey18b,Kipping25}, around longer-period gas giant exoplanets.
These detections are contested: other teams have carried out an analysis on the data and do not find the exomoon signal \citep{Kreidberg19,Heller24}.

Directly imaged exoplanets offer a greater potential for satellite detection due to their typically much larger separations from the host star, which increases the planet's Hill sphere radius \citep{Dobos21}, and the absence of significant inward planet migration keeps any attendant moons after their formation.
Astrometric searches can detect exomoons at larger semimajor distances, and an initial search around the substellar companion HD 206893 B with VLTI/GRAVITY+ has resulted in a tantalising signal \citep{2025arXiv251120091K}.


Radial velocity (RV) searches have looked at exoplanets in the HR~8799 system \citep{Vanderburg21}, HR~7672~B \citep{Ruffio23} and GQ~Lup~B \citep{Horstman24} with no detections.
The closest known self-luminous gas giant exoplanet is $\sim$ 11 \mj{} \bpb{} \citep{Lagrange09-1,Lacour21}, located 19 parsecs away around a young \citep[23 Myr; ][]{Bell15} A-type star and embedded in an edge-on circumstellar disk.
The orbit of $\beta$ Pictoris b is highly inclined, with its orbit seen almost close to edge-on \citep{Lacour21}.
Building on this geometry, \citet{Poon24} propose that the planet has a large obliquity, given the measured projected rotational velocity \citep[$v \sin i = 20$ \kms{};][]{Landman24b}, and suggest that a Neptune-mass exomoon at 40 to 70 planetary radii could be the cause.
RV measurements measure the mass function $M\sin i$ and for exomoons we have no measurements for any orbital inclinations.
Moons in the Solar System are generally coplanar with the equator of their parent planet, while \bpb{} has a relatively high inclination as measured from the rotational broadening of absorption lines within its atmosphere \citep{Snellen14}, so it is not an unreasonable assumption to assume any moons around the planet will also have a high inclination as seen from Earth.

In 2016, the planet's Hill sphere transited the star to within 20\% of the Hill sphere radius, and a photometric monitoring program established that there was no significant circumplanetary dust \citep{Kenworthy21}, concluding that any material had already condensed into satellites around the planet and may well have high inclinations amenable for detection with RV methods.

We present upper limits on our search for exomoons around \bpb{} using CRIRES+ combined with astrometric limits derived from archival astrometric data \citep{Macias26}.
In Sect.~\ref{sect:obs} we detail the observations with the VLT, followed by a periodogram analysis with injection and recovery to determine sensitivity limits.
We discuss these results in Sect.~\ref{sect:discuss} along with our conclusions in Sect.~\ref{sect:conclusions}.

\section{Observations and analysis}\label{sect:obs}

\bpb{} was observed as a service mode program with various epochs between UT 17 Oct 2024 and UT 09 Mar 2025 using VLT/CRIRES+.
We used the 0.2'' slit and the K2166 wavelength setting.
The slit was oriented on the planet location at a 45 degree angle with respect to the line connecting the star and the planet.
This allows us to avoid overexposure from the bright star, while still allowing us to use the starlight for calibration purposes.
Each observation consisted of 24 integrations of 120 seconds in an AAABBBBBBAAA nodding scheme.
We chose this approach to minimize the overhead time from the nodding cycles, while still allowing for background subtraction, which is done using the average of the 3 closest exposures in the other nodding position.
We analyze the exposures from the A and B frames separately, as they can have slightly different wavelength solutions.

%
%

We reduced the spectra with two separate pipelines.
In the first one, the initial data reduction, such as background subtraction, flat fielding, and a first wavelength solution, was done with \textit{pycrires} \citep{stolker2023_pycrires}, which makes use of the official CRIRES+ pipeline \footnote{\url{https://www.eso.org/sci/software/pipelines/crires/crires-pipe-recipes.html}}.
We follow the reduction and analysis steps detailed in \citet{Landman24b} to detect the planet and obtain an estimate of its radial velocity, and give a short summary here: First, we refine the wavelength solution using the telluric lines imprinted on the spectra of the host star.
The stellar contamination at the location of the planet is estimated using low-pass filtering and a Principal Component Analysis (PCA).
We use 10 PCA components to model the stellar and telluric contribution.
We then jointly fit the stellar contribution and planet signal using the likelihood framework developed in \citet{ruffio2021AJ_hr8799_osiris} and adapted to CRIRES+ in \citet{Landman24b}.
In this case, we fix all the parameters of the planet, e.g. the atmospheric parameters and rotational velocity, to the values obtained in \citet{Landman24b}.
This likelihood was calculated for a grid of radial velocity values, which was finally converted to a maximum likelihood estimate and an associated uncertainty.
The mean precision was 320~m~s$^{-1}$.

   \begin{figure*}
    \script{plot_spec_betapicb.py}
        \centering
        \includegraphics[width=1.0\textwidth]{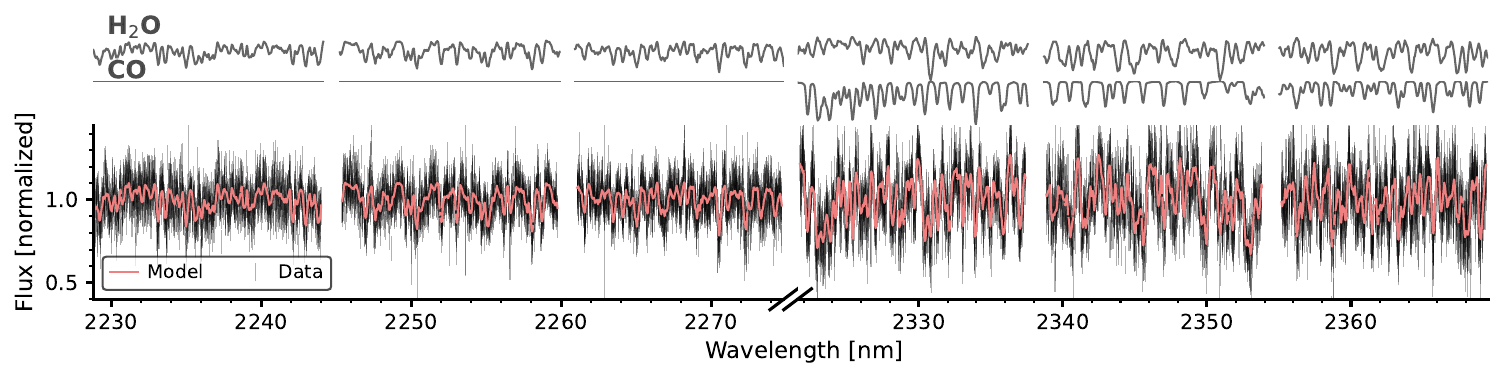}
        {\caption{A section of the spectrum of \bpb{} from CRIRES+, showing one night of data and the fitted model.
        The template spectra of water and carbon monoxide are shown above, indicating their relative contributions.
        }
        \label{fig:criresspec}}%
    \end{figure*}

A second, separate and fully independent reduction and analysis were carried out to validate the RV measurements.
In this workflow, the data were processed using the Python package \texttt{excalibuhr} \citep{Zhang24}.
The reduction steps closely follow those of \texttt{pycrires}; however, we implemented a revised extraction scheme that combines all spatial pixels containing a significant planetary signal.
A section of one of the reduced spectra is shown in Fig.~\ref{fig:criresspec}.
Atmospheric retrievals were then performed using forward models generated with \texttt{petitRADTRANS} \citep{Molliere19} and explored via nested sampling with \texttt{ultranest} \citep{Buchner21}.
This approach allows us to constrain the planet’s atmospheric properties, radial velocity, and rotational broadening.
We adopted state-of-the-art line lists for water \citep[H$_2$O; ][]{polyanskyExoMolMolecularLine2018} and carbon monoxide \citep[CO; ][]{rothmanHITEMPHightemperatureMolecular2010, liROVIBRATIONALLINELISTS2015}.
The atmospheric composition is parameterised using constant-with-altitude abundances, with free parameters for each relevant molecular species, as described in \citet{deregtESOSupJupSurvey2024}.
The temperature structure is modelled using a parametrised profile in which temperature gradients at reference atmospheric layers are retrieved, following the implementation of \citet{gonzalezpicosESOSupJupSurvey2024}.
The complete results and interpretation of the atmospheric retrievals, together with details of the improved spectral extraction scheme, will be provided in a forthcoming paper (Gonzalez Picos et al., submitted).
The contribution of starlight at the planet's location is incorporated directly into the forward model, following the approach demonstrated for GQ Lup B by \citet{GonzalezPicos25}, adapted from \citet{Ruffio23} and \citet{Landman24b}.
We adopt these radial velocities for subsequent analysis due to their greater precision.

   \begin{figure}
        \centering
        \script{plot_cc_betapicb.py}
        \includegraphics[width=1.0\columnwidth]{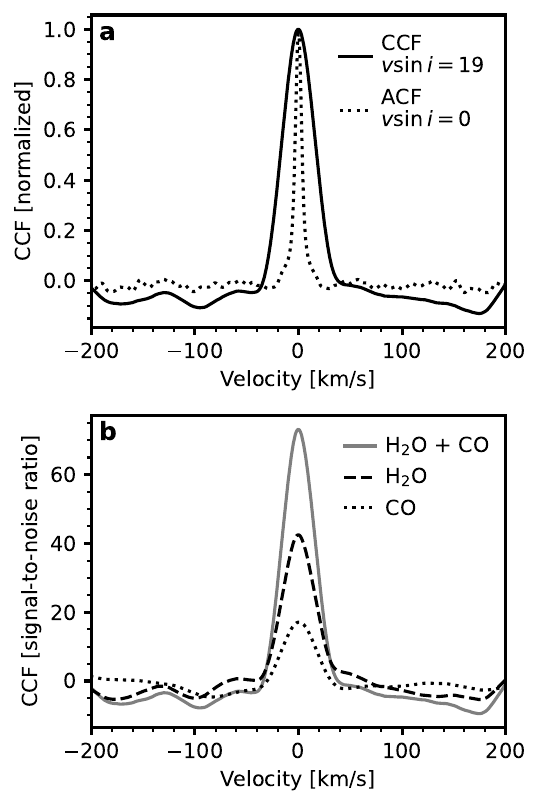}
        {\caption{Cross correlation of the CRIRES+ spectrum of the planet, (a) indicating the rotational broadening of the planet and at the retrieved $v \sin{i} \approx$19 \kms{} and the auto-correlation function (ACF) of the unbroadened model spectrum. (b) the relative detection strengths of the planet in water and carbon monoxide.}
        \label{fig:xcor}}%
    \end{figure}

   \begin{figure}
        \centering
        \includegraphics[width=0.45\textwidth]{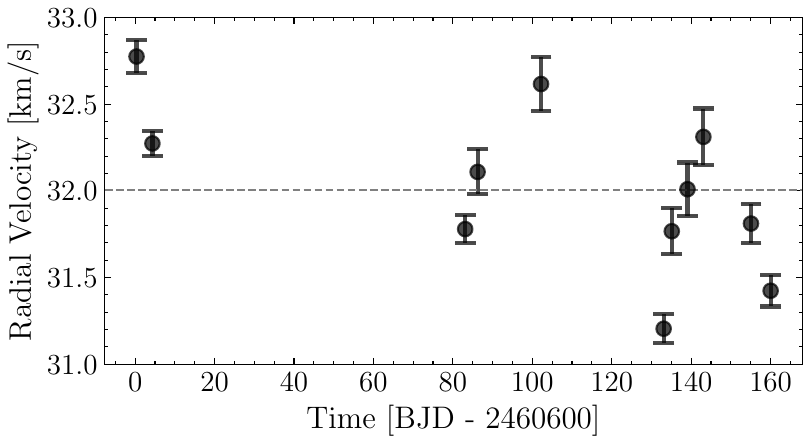}
        {\caption{Radial velocity measurements of \bpb{} from CRIRES+ with the \citet{gonzalezpicosESOSupJupSurvey2024} formulation.}}
        \label{fig:radvel}%
    \end{figure}

In Fig.~\ref{fig:xcor}, we show the detection of the planet's rotation and the identification of water (H$_2$O) and carbon monoxide (CO), which are the dominant opacity sources over the wavelength range of the observations ($\lambda = 2060$--2472 nm).
The planetary spectrum and the best-fit model are shown in Fig.~\ref{fig:criresspec}.
For subsequent analysis we adopt these physical parameters of \bpb{}: from \citet{Lacour21} we take the mass of \bpb{} to be $11.9\pm3.0$\,\mj{}, with orbital parameters $a=9.93\pm0.03$\,au, $e=0.103\pm0.003$ and $i=89.00\pm0.01$ deg.
For the radius we adopt the Exo-REM posterior from \citet{Ravet25} of $1.73\pm0.01$\,\rj{}.
We calculate the Roche limit both for solid and fluid bodies using the mass/radius relationship from \citet{Chen17}, although we note that young objects may have a larger radius and therefore lower density.

\begin{figure}
    \centering
    \includegraphics[width=1.0\linewidth]{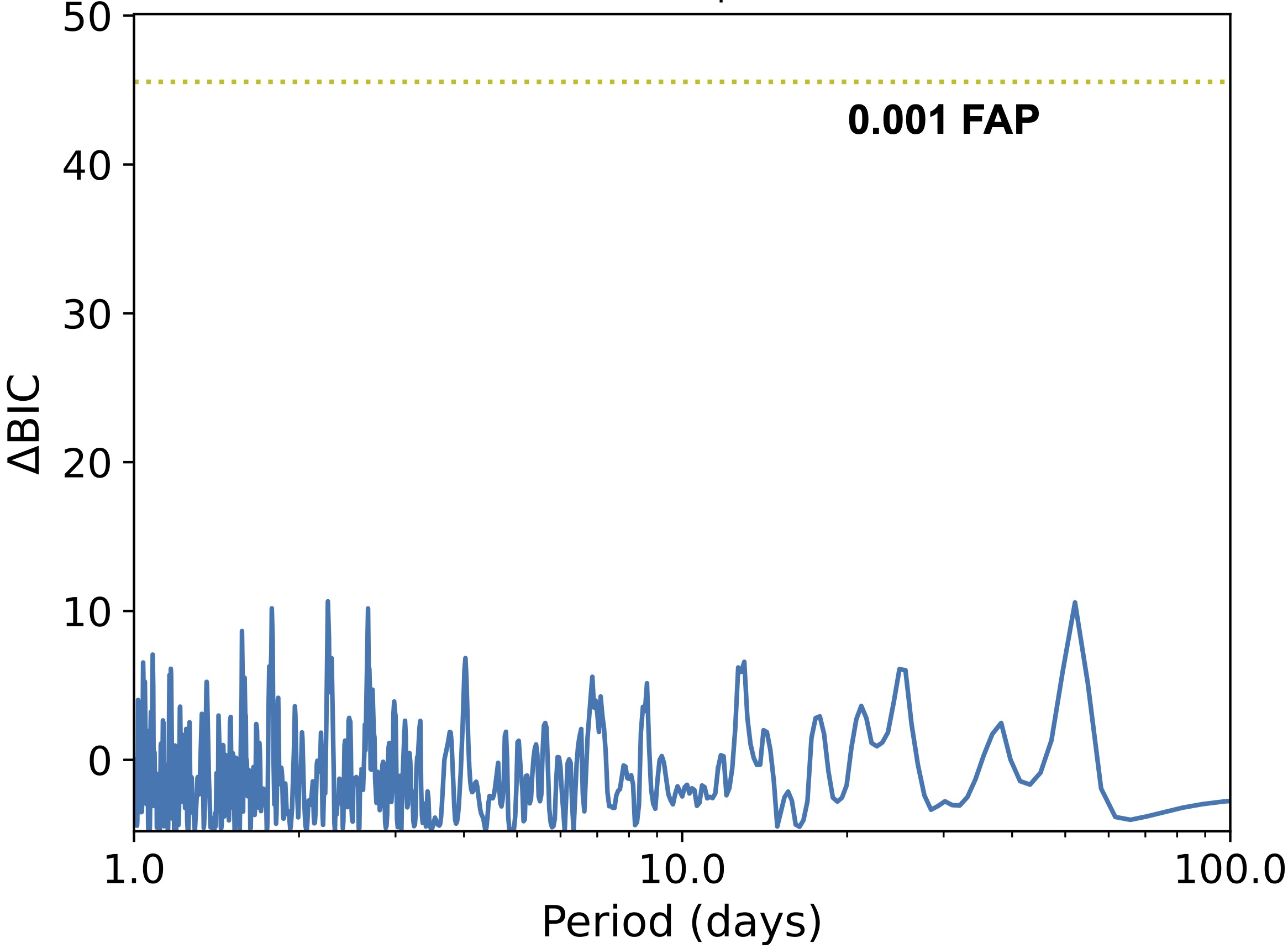}
    \caption{Results of the blind search for exosatellites in the RV time series listed in Table~\ref{tab:rv} using \texttt{RVsearch}. It shows the difference in the Bayesian information criterion ($\Delta$BIC) between a model including a satellite and a model without a satellite. We also show the threshold for the 0.001 false alarm probability (FAP).}
        \label{fig:blindsearch}
\end{figure}



%
%
\begin{table}
\centering                        
   \bgroup
   \caption{Radial velocity measurements of \bpb{}.}    \egroup             

    \def\arraystretch{1.35}
\begin{tabular}{l c c c}      
\hline\hline               
MJD & RV$_{\textrm{instrument}}$ & RV$_{\textrm{corr}}$ & RV$_{\textrm{helio}}$ \\         
 & (\kms{}) & (\kms{}) & (\kms{}) \\
\hline                      
60600.36 & $26.00_{-0.10}^{+0.11}$ & +6.78 & 32.78 \\ 
60604.35 & $25.81_{-0.07}^{+0.07}$ & +6.46 & 32.27 \\ 
60683.10 & $35.17_{-0.08}^{+0.09}$ & -3.39 & 31.78 \\
60686.23 & $36.11_{-0.13}^{+0.11}$ & -4.00 & 32.11 \\
60702.20 & $38.46_{-0.16}^{+0.13}$ & -5.84 & 32.66 \\
60733.07 & $39.13_{-0.08}^{+0.08}$ & -7.92 & 31.20 \\
60735.12 & $39.83_{-0.13}^{+0.13}$ & -8.06 & 31.77 \\
60739.08 & $40.13_{-0.15}^{+0.15}$ & -8.12 & 32.01 \\
60743.08 & $40.51_{-0.16}^{+0.14}$ & -8.20 & 32.31 \\
60755.06 & $39.97_{-0.11}^{+0.08}$ & -8.16 & 31.81 \\
60760.01 & $39.41_{-0.09}^{+0.10}$ & -7.99 & 31.42 \\
\hline                                  
\end{tabular}
\label{tab:rv}    
\end{table}

   \begin{figure*}
        \centering
        \script{plot_rbpic_exomoon_limits.py}
        \includegraphics[width=1.0\textwidth]{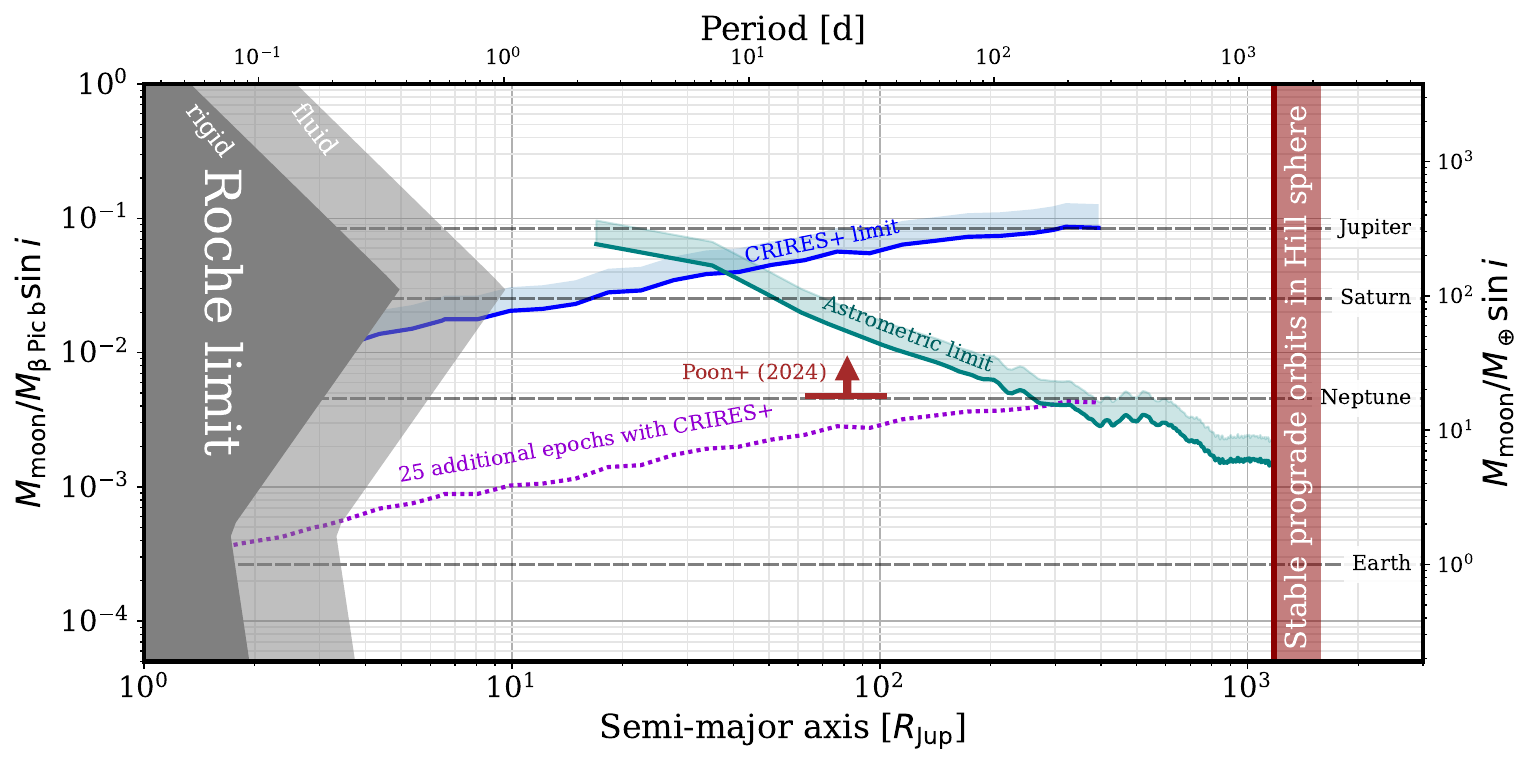}
        {\caption{The upper limits on exosatellites around \bpb{}.
        The blue line represents the radial velocity limits for 50\% retrieval of injected edge-on circular orbits.
        The red line is the outermost stable prograde orbit for an exosatellite around \bpb{}.
        The pale green line corresponds to astrometric limits from \citet{Macias26}.
        The Roche limits for rigid and fluid exosatellite bodies are shown in gray on the left.
        The predicted mass and location for an exomoon which could give Beta Pic b a high obliquity is indicated in purple \citep{Poon24}.
        The predicted sensitivity for an additional 25 observations with CRIRES+ is shown in purple.
        }
        \label{fig:exolimits}}%
    \end{figure*}

%
%

Analysis of the RV measurements was done using the \texttt{RVsearch} Python package \citep{Rosenthal21}.
The algorithm fits both a ``no-moon'' model and ``one moon'' model for a given set of trial periods from 1 to 500 days in a circular orbit, and the difference between these two Bayesian information criterion (BIC) models, $\Delta$BIC, is calculated.
It then estimates a $\Delta$BIC detection threshold that corresponds to a false alarm probability of 0.1\%, following the methodology outlined in \citet{Howard16} where further details can be found.
The results are shown in Fig.~\ref{fig:blindsearch}.
Even though the shortest time between two observations is around two days, we are sensitive to shorter orbital periods due to the irregularly spaced sampling of our measurements.
The largest $\Delta$BIC has a value of 12 for an orbital period of 2.3 days, below the false alarm probability threshold of 46.
A set of injection and recovery tests for various trial periods and masses for a single exomoon on a circular orbit was performed, with the resultant recovery limits shown in Fig.~\ref{fig:sensitivity}.
Fig.~\ref{fig:exolimits} shows the upper limits for a single exomoon orbiting \bpb{} in a circular orbit, as a function of semi-major axis and moon to planet mass function ratio.

\section{Discussion}\label{sect:discuss}

The mass ratio of moons to planets is around $10^{-4}$ for the gas giants \citep{Canup06}, but theoretical studies suggest that more massive planets have more massive moons with a scaling law of $M_{moon} \propto M_P^{3/2}$ with ratios approaching $10^{-3}$ \citep[see Eq. 43 in ][]{Batygin20}.
From the RV analysis, no compelling periodic signal is seen in the data presented in the previous section.
The smallest mass ratio reached is $2\times10^{-2}$ ($60M_\oplus$) at the fluid Roche limit, just below a Saturn mass exomoon with an orbital period of 0.9 days, gradually increasing to a mass ratio of just under $0.1$ at 250 days.
The sensitivity is comparable to previous RV searches around other substellar and planetary companions.
A study on the HR 8799 b,c,d exoplanets with Keck OSIRIS radial velocities ruled out most companions with $m\sin i > 2$\mj{} for orbital periods shorter than 5 days \citep{Vanderburg21} whereas we reach down to just over a Saturn mass for \bpb{}, whilst a limit of $10^{-2}$ (100 $M_\oplus$) was reached for orbital periods of 1 day around the substellar companion GQ Lup B \citep{Horstman24}.
%

%
%

An additional observing campaign, consisting of 25 CRIRES+ observations sampled randomly over a 50 day period, was simulated and added to the current RV dataset, and the resultant sensitivity curve is shown in Fig.~\ref{fig:exolimits} with the injection recovery limits shown in Fig.~\ref{fig:future}.
Given the previous performance of CRIRES+, we can reach a mass ratio of $10^{-3}$ for an orbital period of 1 day, and we can potentially detect Neptune mass exomoons out to periods of 200 days.


\citet{Macias26} analyzed archival astrometric data on \bpb{} and determined upper limits on exomoons. 
These limits are shown as the pale green line in Fig.~\ref{fig:exolimits}.
%
Astrometric detection limits complement the radial velocity methods, and we can now rule out $150 M_\oplus$ exomoons for edge-on circular orbits around \bpb{} for all prograde stable orbits \citep[estimated as 0.44 of the Hill sphere radius using Eq. 5 from ][]{Domingos06}.
Indeed, even though there is a peak in the periodogram at around 52 days (but still below the FAP limit), the astrometric limits are far more constraining at this period and show no signal.  

\section{Conclusions}\label{sect:conclusions}

These reported measurements represent the most sensitive limits to exomoons at short orbital periods to date for \bpb{}, and the complementarity of astrometric limits give coverage throughout the Hill sphere of the planet.
Extending the monitoring of \bpb{} with CRIRES+ will increase the sensitivity to moons with mass ratios greater than about $5\times10^{-3}$ i.e. moons that formed/arrived via gravitational capture pathways or ones which induce the obliquity in \bpb{} \citep{Poon24} - see Fig.~\ref{fig:exolimits}.
Together with the first astrometric limits of \bpb{} from \citet{Macias26} we can rule out any exomoons three times more massive than Saturn in the Hill sphere.

%
%

The discovery of massive gas giant exoplanets at large separations from their parent stars, such as HD106096~b \citep{Bailey14}, YSES-1 b and c \citep{Bohn20} and WISPIT~1 b and c \citep{vanCapelleveen25b} enable direct spectroscopic monitoring of the planets to look for exosatellites.
One highly promising candidate is YSES~1b which hosts both a circumplanetary disk seen in thermal emission \citep{Hoch25} and whose orbit is almost edge on to our line of sight \citep{Roberts25}.
If the geometry is particularly favourable and the moon transits the disk of the exoplanet, the Rossiter-McLaughlin effect can produce a very pronounced RV signal due to the rapid rotation of these young exoplanets \citep{Ruffio23}.
If exomoons do transit \bpb{} and other wide projected separation planets such as YSES~1~b, then we may be highly fortunate and have our first RV-detected exomoons within the next few years.

\section*{Acknowledgements}

We thank our referee for their careful reading of our manuscript and suggestions which improved it.
JLB acknowledges funding from the European Research Council (ERC) under the European Union’s Horizon 2020 research and innovation program under grant agreement No 805445, and the support of the Leverhulme Trust via the Philip Leverhulme Physics Prize.
Based on observations collected at the European Southern Observatory under ESO programme 114.27DX.001.
This research has used the SIMBAD database, operated at CDS, Strasbourg, France \citep{Wenger00}.
This work has used data from the European Space Agency (ESA) mission {\it Gaia} (\url{https://www.cosmos.esa.int/gaia}), processed by the {\it Gaia} Data Processing and Analysis Consortium (DPAC, \url{https://www.cosmos.esa.int/web/gaia/dpac/consortium}).
Funding for the DPAC has been provided by national institutions, in particular the institutions participating in the {\it Gaia} Multilateral Agreement.
To achieve the scientific results presented in this article, we made use of the \emph{Python} programming language\footnote{Python Software Foundation, \url{https://www.python.org/}}, especially the \emph{SciPy} \citep{Virtanen20}, \emph{NumPy} \citep{Oliphant07}, \emph{Matplotlib} \citep{Hunter07}, \emph{emcee} \citep{ForemanMackey13}, \emph{astropy} \citep{astropy:2013,astropy:2018} packages.
%

\section*{Data Availability}

 This paper is compiled using the showyourwork! framework, which includes all original data files, computer scripts to generate the figures, and associated compilation framework.
 This is available at \url{https://github.com/mkenworthy/BetaPicbExomoons}

\bibliographystyle{mnras}
\bibliography{bib} 


\appendix

\section{Injection recovery tests}
\label{app:injrecover}

   \begin{figure}
        \centering
        \includegraphics[width=0.5\textwidth]{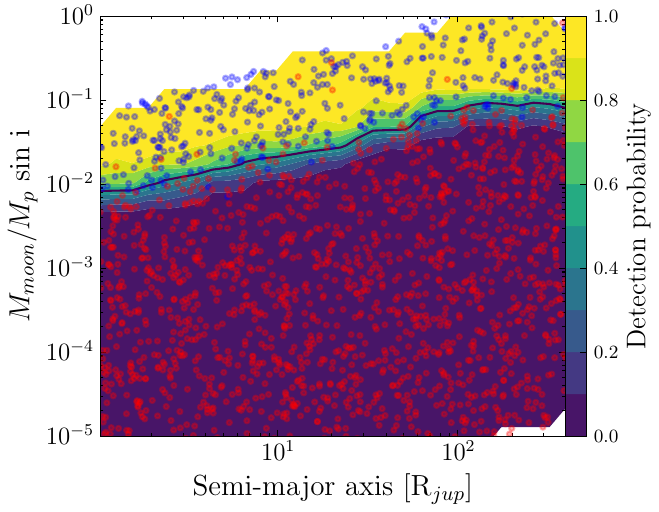}
        {\caption{Exosatellite detection limits derived using injection and recovery tests for \bpb{}.}
        \label{fig:sensitivity}}
    \end{figure}

   \begin{figure}
        \centering
        \includegraphics[width=0.5\textwidth]{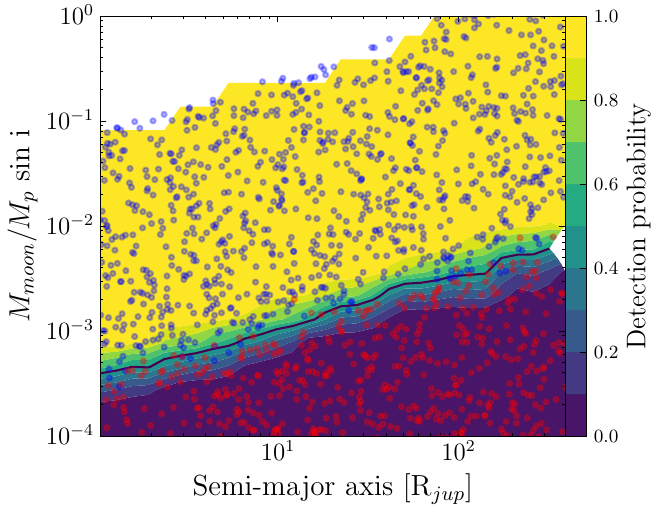}
        {\caption{Exosatellite detection limits with 25 additional epochs of CRIRES+ observations with precision of 250 \ms{} in the 2025/26 observing season.}
        \label{fig:future}}
    \end{figure}


\bsp	
\label{lastpage}
\end{document}